\documentclass[prd,aps,tightline,twocolumn,nofootinbib,superscriptaddress]{revtex4}

\usepackage{graphicx}
\usepackage{amssymb}
\usepackage{amsmath}
\usepackage{color}

\begin{document}

\newcommand{\be}{\begin{equation}}
\newcommand{\ee}{\end{equation}}







\title{
Cosmological aspects of inflation in a supersymmetric axion model
}




\author{Masahiro Kawasaki}
\affiliation{Institute for Cosmic Ray Research, University of Tokyo, Kashiwa, Chiba 277-8582, Japan}
\affiliation{Institute for the Physics and Mathematics of the Universe, University of Tokyo, Kashiwa, Chiba 277-8568, Japan}
\author{Naoya Kitajima}
\affiliation{Institute for Cosmic Ray Research, University of Tokyo, Kashiwa, Chiba 277-8582, Japan}
\author{Kazunori Nakayama}
\affiliation{Department of Physics, University of Tokyo, Bunkyo-ku, Tokyo 113-0033, Japan}
\affiliation{KEK Theory Center, Institute of Particle and Nuclear Studies, KEK, Tsukuba, Ibaraki 305-0801, Japan}

\date{\today}

\vskip 1.0cm

\begin{abstract}
We show that the hybrid inflation is naturally realized 
in the framework of a supersymmetric axion model,
which is consistent with the WMAP observation 
if the Peccei-Quinn symmetry breaking scale is around $10^{15}$GeV.
By solving the post-inflationary scalar dynamics, it is found that 
the scalar partner of the axion, saxion, oscillates with large amplitude and 
its decay produces a huge entropy and dilutes the axion.
As a result, the axion coherent oscillation can be the dominant component of the dark matter
in the Universe. Cosmological gravitino and axino problems are solved.
\end{abstract}


\maketitle

\section{Introduction}

The standard model of particle physics has succeeded in explaining 
many experimental facts until now, 
but there exist several problems which have not been solved yet.
One of them is the strong CP problem.
The Lagrangian of quantum chromodynamics (QCD) allows the 
CP-violating term, but the neutron electric dipole measurement implies that
QCD preserves CP and hence the CP-violating term is stringently constrained. 
The most popular solution for the problem was proposed by 
Peccei and Quinn~\cite{Peccei:1977hh}.
They introduced an additional global $U(1)$ symmetry, called 
Peccei-Quinn (PQ) symmetry, written as $U(1)_\mathrm{PQ}$.
When the PQ symmetry is broken spontaneously, a pseudo-Nambu-Goldstone 
boson appears, which is called an axion.
The axion acquires its mass through the QCD instanton effect and the axion 
settles down to the potential minimum ($=$ vacuum) where CP is preserved~\cite{Kim:1986ax}. 

Another well-known problem of the standard model is the gauge hierarchy problem.
The electroweak scale is unstable against the radiative correction. 
The squared Higgs mass receives a quadratically divergent correction
which is thought to be, say, the grand unification scale $\sim 10^{16}$~GeV. 
Thus keeping the Higgs mass around the weak scale requires unnatural fine-tuning. 
This problem is solved if we introduce the supersymmetry (SUSY) 
which is the symmetry with respect to the replacement of bosons with 
fermions. In SUSY theory, the quadratically divergent quantum correction is 
canceled~\cite{Martin:1997ns}.
Thus we have a good motivation for considering the axion model 
in the framework of SUSY.

On the other hand,
the cosmological observations support that the Universe underwent 
an accelerated expansion, called inflation, 
at the very early stage of the Universe~\cite{Komatsu:2010fb}.
It is generally considered that inflation is driven by potential energy of a scalar field, inflaton.
Thanks to inflation, problems of the standard cosmology such as the 
horizon problem and the monopole problem are solved.
In addition, inflation can produce the density perturbation which 
accounts for the large scale structure of the present Universe.
The observation by the Wilkinson Microwave Anisotropy Probe (WMAP) satellite 
confirmed the nearly scale invariant density perturbation, which is a generic prediction of inflation.

The cosmological observations also revealed the existence of the nonbaryonic 
dark matter (DM), which occupies 23\% of the total energy of the present Universe~\cite{Komatsu:2010fb}.
The standard model (SM) in particle physics does not contain candidates for DM,
and hence we need the physics beyond the SM in order to explain the existence of DM.

In the previous paper~\cite{Kawasaki:2010gv}, 
we pointed out that a SUSY axion model naturally
causes hybrid inflation and the axion becomes the dominant component of DM.
In the hybrid inflation model, the PQ scalar fields play a role of 
the waterfall field and the PQ phase transition takes place at the end of inflation.
To account for the amplitude of the density perturbation observed 
by the WMAP, the PQ scale $f_a$ must be of order of $10^{15}$~GeV.
This value seems to be too large because there is an upper bound  
$f_a \lesssim 10^{12}$~GeV in order for the axion not to overclose the Universe.
However, in the SUSY axion model it is found that the axion is diluted 
by the late-time entropy production via saxion decay and the upper 
bound is relaxed to $10^{15}$~GeV~\cite{Steinhardt:1983ia, Kawasaki:1995vt}, 
which is consistent with the condition for the correct density perturbation.
In this paper we further describe the details of the scalar field dynamics
after inflation in this model by performing detailed numerical calculations and we show that the saxion
field deviates from the true minimum by the finite-temperature effect
and the saxion oscillation has a large amplitude, which leads to a huge 
entropy production and dilutes the axion at a later epoch. 
We also study several problems in our model such as the axion domain wall and baryogenesis
and propose several solutions to these problems. 

This paper is organized as follows.
In Sec. 2 we briefly introduce a SUSY axion model.
In Sec. 3 we review the SUSY hybrid inflation model and show the consistency with observation.
Particularly we point out that the superpotential of the SUSY hybrid inflation model has the 
same form as that of the SUSY axion model, and the PQ scale is determined to be $\sim 10^{15}$GeV
for the PQ sector to induce a hybrid inflation and correctly reproduce observed density perturbation.
In Sec. 4 the dynamics after inflation is discussed and we see that the saxion oscillation with large initial amplitude is induced.
In Sec. 5 it is shown that the late-time entropy production due to the saxion decay necessarily takes place,
and as a result, the axion coherent oscillation can be the dominant component of DM.
In Sec. 6 we discuss the fate of topological defects, such as axionic strings and domain walls.
In Sec. 7 we present a mechanism to create a correct amount of baryon asymmetry under the late-time entropy production.
In Sec. 8 a variant type of SUSY axion model is presented, which causes a so-called smooth-hybrid inflation
and we describe some cosmological aspects of the model.
We conclude in Sec. 9.

\section{Supersymmetric axion model}

\subsection{The potential of the SUSY axion model}

Let us describe the SUSY axion model. 
Here we assume the gravity-mediated SUSY breaking.
The superpotential for the SUSY axion model is given by 
\be
	W = \kappa S (\Psi \bar\Psi - f_a^2) 
	+ \lambda \Psi X \bar X ,
	\label{susy_axion_superpot}
\ee
where, $S$ is a gauge singlet superfield and has a zero PQ charge, 
and $\Psi$ and $\bar\Psi$ are the PQ superfields that are gauge 
singlets and have $+1$ and $-1$ PQ charges, respectively.
The PQ fields contain the axion ($a$), saxion ($\sigma$, the scalar partner of the axion), and 
axino ($\tilde a$, the fermionic superpartner of the axion).
Here $f_a$ is the PQ symmetry-breaking scale and $\kappa$ is a 
dimensionless coupling constant assumed to be real and positive.
$X (\bar X)$ is the superfield interacting with a PQ field
at tree level and has some PQ charges as well as gauge charges
through which it interacts with the minimal supersymmetric standard model (MSSM) fields.
The superpotential also has an $R$-symmetry.
The charge assignments of the fields in the present model are shown in Table~\ref{charge_asignments}.
In particular, for the Kim-Shifman-Vainshtein-Zakharov (KSVZ) (or hadronic) axion model~\cite{Kim:1979if},
$X$ and $\bar X$ are additional heavy quarks, denoted by $Q$ and $\bar Q$, that have color charges.
For the Dine-Fischler-Srednicki-Zhitnitsky (DFSZ) axion model~\cite{Dine:1981rt}, $X$ and $\bar X$ are identified as MSSM Higgses, $H_u$ and $H_d$.\footnote{
	In the DFSZ model, the coupling constant $\lambda$ must be very small,
	say, $\lambda \sim 10^{-12}$ for $f_a \sim 10^{15}$GeV, in order to produce a sizable $\mu$-term.
	This might be a tuning, but it is relaxed by changing the relative PQ charge assignments 
	between $\Psi (\bar\Psi)$ and $H_u (H_d)$.
	For example, if the PQ charges of $H_u$ and $H_d$ are $-n$, where $n (\geq 1)$ is a positive integer,
	the allowed term in the superpotential is
	$\lambda \Psi^{2n} H_u H_d / M^{2n-1}$ with some cutoff scale $M$, which might be the Planck scale.
	In this case the amount of tuning for the coupling constant $\lambda$ is relaxed.
	The phenomenology discussed in the following sections is not modified by the choice of PQ charges for $H_u$ and $H_d$.
}

According to the superpotential~(\ref{susy_axion_superpot}), 
the $F$-term scalar potential is derived as
\be
	V_F = \kappa^2 |\Psi \bar\Psi - f_a^2|^2 
	+ \kappa^2 |S|^2 ( |\Psi|^2 + |\bar\Psi|^2 ),
\ee
where we denote the scalar fields using the same symbol as 
the superfields and we set $X$ and $\bar X$ to be zero assuming that 
they have large Hubble masses and quickly settle down to zero during inflation.
The global minimum of this potential is located at $S=0$ and $\Psi \bar\Psi = f_a^2$.
Here it should be noted that there exists a flat direction 
along which the scalar fields do not feel the potential, 
ensured by the U(1)$_{\rm PQ}$ symmetry extended to a complex U(1) 
due to the holomorphy of the superpotential~\cite{Kugo:1983ma}.
The flat direction is lifted up by the SUSY-breaking effect,  
leading to the following soft SUSY-breaking mass terms :
\be
	V_\mathrm{soft} = c_1 m_{3/2}^2 |\Psi|^2 
	+ c_2 m_{3/2}^2 |\bar\Psi|^2 ,
\ee
where $m_{3/2}$ is the gravitino mass and $c_1$ and $c_2$ 
are real-valued constants that are positive and order unities.
With this soft SUSY-breaking potential, the radial 
components of the PQ fields $|\Psi|$ and $|\bar\Psi|$ are stabilized at 
\be
	v \simeq \bigg( \frac{c_2}{c_1} \bigg)^{1/4} f_a ,
	\quad 
	\bar v \simeq \bigg( \frac{c_1}{c_2} \bigg)^{1/4} f_a,
	\label{vev_of_PQ}
\ee
respectively.\footnote{
	See Refs.~\cite{Asaka:1998ns,Abe:2001cg,Banks:2002sd,Nakamura:2008ey,Carpenter:2009sw,Jeong:2011xu,Higaki:2011bz}
	for other types of the saxion stabilization mechanisms and their cosmological issues.
} 
The saxion field $\sigma$ is defined by the deviation of $|\Psi|$ 
from the vacuum expectation value (\ref{vev_of_PQ}) along the flat direction.

Near the vacuum expectation values (\ref{vev_of_PQ}), 
the axion $a$ and saxion $\sigma$ are related to the PQ fields as
\be
	\Psi = v \exp \bigg[ \frac{\sigma + ia}{\sqrt 2 F_a} \bigg] ,
	\quad 
	\bar\Psi = \bar v \exp \bigg[ -\frac{\sigma + ia}{\sqrt 2 F_a} 
	\bigg], 
\ee
where $F_a$ is determined by requiring that $\sigma$ and $a$ are canonically normalized and given by
$F_a \equiv \sqrt{v^2 +\bar v^2}$.

\begin{table}[tb]
\caption{Charge assignments on the field content}
\begin{ruledtabular}
	\begin{tabular}{cccccc}
		& $S$ & $\Psi$ & $\bar\Psi$ & $X$ & $\bar X$ \\ \hline
		$U(1)_{\mathrm{PQ}}$ & 0 & $+1$ & $-1$ & $-1/2$ & $-1/2$ \\ 
		$U(1)_R$ & +2 & 0 & 0 & +1 & +1 \\ 
	\end{tabular}
\end{ruledtabular}
\label{charge_asignments}
\end{table}

\subsection{The decay of the saxion}

In this subsection, we derive the decay rate of the saxion which is important in the later section.
The kinetic terms of the PQ fields lead the interaction of the saxion 
with the axion~\cite{Chun:1995hc,Kawasaki:2007mk} as 
\be
	|\partial_\mu \Psi|^2 + |\partial_\mu \bar\Psi|^2 
	= \bigg( 1 + \frac{\sqrt 2 \xi}{F_a} \sigma \bigg) 
	\bigg( \frac{1}{2} (\partial_\mu a)^2 
	+ \frac{1}{2} (\partial_\mu \sigma )^2 \bigg) + \cdots,
	\label{eq:saxion_axion_int}
\ee
where $\xi$ is defined as $\xi \equiv ( v^2 - \bar v^2) / F_a^2$, 
which is generally of order unity unless $v \simeq \bar v$ i.e. 
$c_1 \simeq c_2$.
From (\ref{eq:saxion_axion_int}), the decay rate of the saxion 
into two axions is estimated as 
\be
	\Gamma_{\sigma \to aa} = 
	\frac{\xi^2}{64 \pi} \frac{m_\sigma^3}{F_a^2}.
	\label{eq:decay_rate_two_axion}
\ee
Note that the decay rate into two axions can be suppressed by tuning 
$c_1$ and $c_2$ as $ c_1 \simeq c_2$.

In the KSVZ axion model, the dominant decay mode into 
the SM particles is that into two gluons.
The decay rate is calculated as
\be
	\Gamma_{\sigma \to gg} = 
	\frac{\alpha_s^2}{32 \pi^3} \frac{m_\sigma^3}{F_a^2}, 	
	\label{eq:decay_rate_KSVZ}
\ee
where $m_\sigma$ is the mass of the saxion and $\alpha_s$ is 
the QCD gauge coupling constant.
Comparing (\ref{eq:decay_rate_KSVZ}) with (\ref{eq:decay_rate_two_axion}),
the dominant decay mode of the saxion is generally the two-axion decay 
in the KSVZ axion model unless $c_1 \simeq c_2$.
In a minimal setup where the SUSY-breaking masses for $\Psi$ and $\bar\Psi$ only come from
the supergravity effects, $c_1$ is equal to $c_2$ at tree level.
The radiative correction through heavy quarks, which is relevant only for $\Psi$,
produces a difference between $c_1$ and $c_2$ at the low energy scale~\cite{Chun:1995hc}.
However, even if such effects are taken into account, the two-gluon decay can be the dominant mode.

In the DFSZ axion model, there exists a tree level coupling of 
the saxion with the standard model Higgses, as
$\mathcal L_\mathrm{int} = \lambda^2 \sigma^2 
(|H_u^0|^2 + |H_d^0|^2)$. Thus the saxion decays into Higgses
with decay rate given by
\be
	\Gamma_{\sigma \to hh}=
	\frac{1}{8\pi} \frac{m_\sigma^3}{f_a^2} 
	\left( \frac{\mu}{m_\sigma} \right)^4
	\left( 1-\frac{4m_h^2}{m_\sigma^2} \right)^{1/2}, 		
	\label{decay_rate_DFSZ}
\ee
where $\mu = \lambda \langle \Psi \rangle$ gives the Higgsino mass.
In the DFSZ axion model, this decay mode may dominate over the two-axion decay.

\section{Hybrid inflation in a SUSY axion model}

In this section, we show that the axion model described in the 
previous section is also considered as a SUSY hybrid 
inflation model~\cite{Copeland:1994vg,Nakayama:2010xf}.\footnote{
	The different SUSY hybrid inflation models related to PQ symmetry 
	are found in~\cite{BasteroGil:1997vn,Lazarides:2010di}.
}
The relevant parts of the  superpotential and K\"ahler potential
which are responsible for inflation are written as
\begin{align}
	K & = |S|^2 + |\Psi|^2 + |\bar\Psi|^2 
	+ k_S \frac{|S|^4}{4M_P^2} \notag \\
	& + k_1 \frac{|S|^2 |\Psi|^2}{M_P^2} 
	+ k_2 \frac{|S|^2 |\bar\Psi|^2}{M_P^2}
	+ k_{SS} \frac{|S|^6}{6M_P^4} + \dots , \\[2mm]
	W & = \kappa S (\Psi \bar\Psi - f_a^2)+W_0, 
	\label{superpot}
\end{align}
where $S$ and $\Psi (\bar\Psi)$ take roles of the inflaton and 
waterfall fields, respectively; $W_0 = m_{3/2}M_P^2$ is the constant term which cancels the 
vacuum energy coming from the SUSY-breaking effect and makes the cosmological constant nearly zero
in the present Universe;
and $k_S$, $k_1$, $k_2$, and $k_{SS}$ are dimensionless real-valued coefficients.
As seen later, the PQ scale $f_a$ sets the energy scale of inflation.

In the framework of the supergravity, the scalar 
potential is obtained from 
\be
	V = e^{K / M_P^2} \bigg[ K^{ij^*} D_iW D_j^* W^* 
	- 3\frac{|W|^2}{M_P^2} \bigg]
\ee
where $M_P$ is the Planck scale, $D_i W = W_i + K_i W / M_P^2$, $K^{ij^*} = K^{-1}_{ij^*}$, and 
subscript $i$ means the derivative with respect to the field 
$\phi_i = \{ S, \Psi, \bar\Psi\}$.
In the global SUSY limit, i.e. $M_P \to \infty$, 
the $F$-term potential that comes from the superpotential (\ref{superpot}) 
becomes
\be
	V = \kappa^2 | \Psi \bar\Psi - f_a^2 |^2 
	+ \kappa^2 |S|^2 ( |\Psi|^2 + |\bar\Psi|^2 ). 
	\label{F-term_pot}
\ee
The minimum of the potential lies at $S = 0$ and 
$\Psi \bar\Psi = f_a^2$, where $V=0$.
However, if $|S|$ takes a large value, the (local) minimum appears 
at $ \Psi = \bar\Psi = 0$ where $V=\kappa^2 f_a^4$.
Since the potential for $|S| \gg f_a$ with $\Psi = \bar\Psi = 0$ is flat, inflation takes place there. 
Once $|S|$ rolls down from a larger value and reaches $|S|=f_a$, 
the waterfall behavior turns on and the inflation ends suddenly.
To see this, we rewrite the mass terms for $\Psi$ and $\bar\Psi$ in the potential (\ref{F-term_pot}) as
\be
	V = \begin{pmatrix} \Psi^* & \bar\Psi \end{pmatrix}
	\begin{pmatrix} \kappa^2 |S|^2 & -\kappa^2 f_a^2 \\[2mm] 
	- \kappa^2 f_a^2 & \kappa^2 |S|^2 \end{pmatrix}
	\begin{pmatrix} \Psi \\[2mm] \bar\Psi^* \end{pmatrix} 
	+ \cdots .
\ee
So there are two mass eigenstate with squared masses  
$\kappa^2 ( |S|^2 \pm f_a^2)$.
Therefore when $|S| < f_a$, one of the mass eigenstates becomes 
tachyonic and gets a nonzero vacuum expectation value.
After that, $S$ approaches to zero and $\Psi$ and $\bar\Psi$
relax to the flat direction given by $\Psi \bar\Psi = f_a^2$. 

We rewrite the inflaton $S$ as $S \equiv \varphi e^{i\theta} /\sqrt{2} $.
During the inflation period, including supergravity effect, 
the scalar potential is given by 
\be
	\begin{split}
	V = \kappa^2 f_a^4 \bigg( 1 - k_S \frac{\varphi^2}{2M_P^2} 
	+ \gamma_S \frac{\varphi^4}{8M_P^4} + \cdots \bigg) \\[1mm]
	+2 \kappa f_a^2 m_{3/2}(S+S^*)
	+ \Delta V_\mathrm{1-loop},
	\label{inflaton_pot}
	\end{split}
\ee
where $\gamma_S \equiv 1 - 7 k_S / 2 + 2 k_S^2 - 3 k_{SS}$ and
$\Delta V_\mathrm{1-loop}$ is the one-loop radiative correction. 
Note that the squared masses of the scalars and fermions in 
superfields $S$, $\Psi$, and $\bar{\Psi}$ are
\begin{align}
	& ~~~~ S     ~~~~~~~~~~~~~ \Psi,~\bar{\Psi} 
	\nonumber \\[2mm]
	\text{scalar : }& 0 \times 2,\quad \kappa^2 ( |S|^2 \pm M^2 ) 
	\times 2 \\[2mm]
	\text{fermion : }& 0 \times 2, \quad \kappa^2 |S|^2 \times 4.
\end{align}
Then, the one-loop radiative correction in the scalar 
potential~\cite{Coleman:1973jx} is calculated as 
\be
	\Delta V_\mathrm{1-loop} = 
	\sum_i (-1)^F \frac{m_i^4}{64\pi^2} \ln \frac{m_i^2}{\Lambda^2} 
	= \frac{\kappa^4 f_a^4}{8\pi^2} F(x),
\ee
where the sum is taken over the field degrees of freedom and $F=0$ for scalar and $F=1$ for fermion.
Here $\Lambda$ is some cutoff scale,
$x$ is defined as $x \equiv |S| / f_a = \varphi / \sqrt 2 f_a$, and
\be
	\begin{split}
	F(x) \equiv \frac{1}{4} \bigg[ & (x^4 + 1) \ln \frac{x^4-1}{x^4}  \\[1mm]
	& + 2x^2 \ln \frac{x^2 + 1}{x^2 - 1} 
	+ 2 \ln \frac{\kappa^2 M^2 x^2}{\Lambda^2} -3 \bigg].
	\end{split}
\ee
In particular, for $x \gg 1$, i.e. $\varphi \gg \sqrt 2 f_a$, 
the one-loop correction can be approximated as 
\be
	\Delta V_\mathrm{1-loop} \approx \frac{\kappa^4 f_a^4}{16 \pi^2} 
	\ln \frac{\kappa^2 \varphi^2}{2 \Lambda^2}.
\ee

From the scalar potential, the slow-roll parameters~\cite{Liddle&Lyth} are derived as
\begin{widetext}
\begin{equation}
	\epsilon \equiv  \frac{M_P^2}{2}\left( \frac{V'}{V} \right)^2
	= \left( -k_S \frac{\varphi}{\sqrt 2 M_P} 
	+ \gamma_S \frac{\varphi^3}{2 \sqrt 2 M_P^3}
	+ \frac{\kappa^2M_P}{16\pi^2 f_a} F'(x) 
	+ \frac{2m_{3/2}M_P \cos \theta}{\kappa f_a^2}
	\right)^2,
\end{equation}
\begin{equation}	
	\eta \equiv M_P^2 \frac{V''}{V} = -k_S + 3\gamma_S \frac{\varphi^2}{2 M_P^2} 
	+ \frac{\kappa^2M_P^2}{16 \pi^2 f_a^2} + F''(x),
\end{equation}
\end{widetext}
where the prime denotes the derivative with respect to $\varphi$.
The number of e-foldings is calculated from
\be
	N = \int ^{t}_{t_f} H dt 
	= \frac{1}{M_P^2} \int^{\varphi}_{\varphi_f} \frac{V(\varphi)}{V'(\varphi)}d\varphi ,
\ee
where the subscript $f$ means the moment when the inflation ends. 
The number of e-foldings after a comoving scale $k_0$ leaves the horizon is written as
\be
	\begin{split}
	N_e = & 56 + \ln \bigg(\frac{ 0.002 \, \mathrm{Mpc^{-1}}}{k_0} \bigg) \\[1mm]
	& + \frac{1}{3} \ln \bigg( \frac{T_R}{10^{10} \, \mathrm{GeV}} \bigg)
	+ \frac{1}{3} \ln \bigg( \frac{H_I}{10^{13} \, \mathrm{GeV}} \bigg),
	\end{split}
\ee
where $T_R$ is the reheating temperature and $H_I$ is the Hubble 
constant during inflation.
Using these quantities, the power spectrum 
of the curvature perturbation $P_\zeta$ and the scalar spectral index 
$n_s$, can be calculated as follows:
\be
	P_\zeta \simeq \frac{1}{12 \pi^2 M_P^6} \frac{V^3}{V'^2},
	\quad n_s \simeq 1 - 6 \epsilon + 2 \eta.
\ee
We can compare these quantities with the results of the WMAP observation~\cite{Komatsu:2010fb},
\be
	P_\zeta = \big( 2.441^{+0.088}_{-0.092} \big) \times 10^{-9},\quad n_s = 0.963 \pm 0.012.
\ee

According to~\cite{Nakayama:2010xf}, to explain the correct magnitude of the density perturbation, 
$f_a \sim 10^{15} \, \mathrm{GeV}$ is needed and by adjusting the coupling constant 
$\kappa$ and the coefficient of a nonminimal K\"ahler term $k_S$
as $\kappa \sim 10^{-3}$ and $k_S \sim 10^{-2}$, 
the scalar spectral index fits well with the WMAP observation.

\section{The post-inflationary dynamics}    \label{sec:post}

In the previous section, we show that the successful inflation 
occurs in the SUSY axion model if the PQ scale is around $10^{15}$GeV.
This value seems to be too large since $f_a \lesssim 10^{12}$~GeV is 
needed in order for the axion not to overclose the Universe.
Note that it is not allowed to tune the amplitude of the axion in the present model,
since the PQ symmetry is restored during inflation and
the misalignment angle takes random values for each small patch of the Universe after inflation.
However, if the late-time entropy production takes place, 
the axion can be diluted so that the bound can be evaded~\cite{Kawasaki:1995vt}.
In this and the next section, we discuss the dynamics of the PQ scalar fields 
after inflation and show that the saxion starts to oscillate with large initial amplitude
and it results in a huge entropy production through its decay.

\begin{figure}[t]
\centering
\includegraphics [width = 8.5cm, clip]{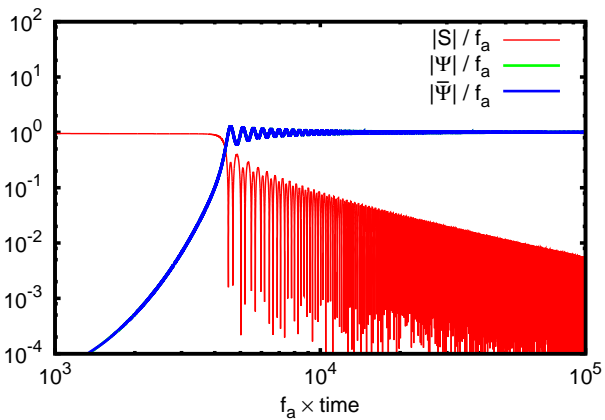}
\caption{
	Time evolution of $S$ (thin red line), $\Psi$ (thick green line), and $\bar{\Psi}$ (thick blue line)
	 after inflation, as a function of time.
	 We have taken $f_a = 0.1M_P, \kappa = 0.01$.
	(The green line and the blue one coincide.)
}
\label{PQ_dynamics1_fig}
\end{figure}

Soon after inflation ends, the inflaton $S$ begins to oscillate around 
its vacuum expectation value (VEV), $S \simeq 0$.\footnote{
	Precisely speaking, the VEV of $S$ deviates from zero because of the supergravity effect.
	The second term in (\ref{inflaton_pot}) makes $\langle S \rangle \sim m_{3/2}/\kappa$ at the true minimum.
}
Since the inflaton oscillation behaves as matter, its amplitude 
decreases inversely proportional to the cosmic time, i.e.  
$\propto 1/t \propto a^{-3/2}$.
The waterfall fields quickly roll down to the flat direction, $\Psi \bar\Psi = f_a^2$, 
as shown in~Fig.\ref{PQ_dynamics1_fig}.
In fact, the flat direction is not flat at this stage,
because both PQ fields $\Psi \bar\Psi$ obtain masses of $\kappa |S|$ before $S$ decays.
Thus, the PQ fields are stabilized at $\Psi= \bar\Psi=f_a$ at this stage
and both fields oscillate around $f_a$.

\begin{figure}
\centering
\includegraphics [width = 8.5cm, clip]{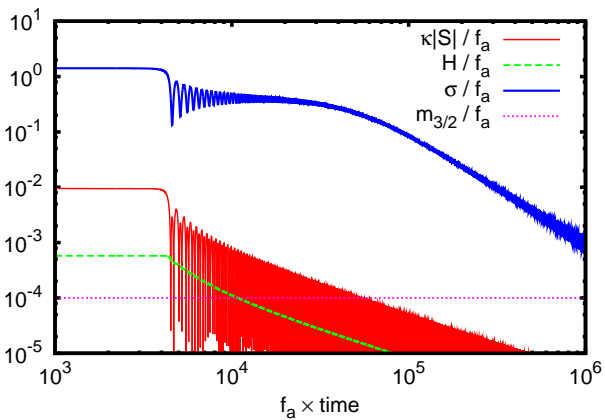}
\caption{
	Evolution of the saxion $\sigma$, inflaton $S$, and Hubble parameter $H$.
	We have taken $f_a = 0.1M_P, \kappa = 0.01$ and $c_1=1, c_2=4$.
	}
\label{fig:saxion_dynamics}
\end{figure}

The evolution of the inflaton and the saxion is shown 
in Fig.~\ref{fig:saxion_dynamics}.
After inflation, PQ scalars oscillate around $|\Psi|=|\bar\Psi|=f_a$ because they have masses of $\kappa |S|$.
When the gravitino mass exceeds $\kappa |S|$, 
the saxion begins to move toward the true minimum (\ref{vev_of_PQ}).
However, the Hubble parameter is already smaller than the gravitino mass
and hence the friction is not efficient.
As a result, the saxion adiabatically approaches to the true minimum without oscillation.
A key point is the existence of the mass of $\kappa |S|$ for PQ scalars, 
which is larger than the Hubble scale.
Thus the saxion oscillation is not induced~\cite{Linde:1996cx}, 
and hence no late-time entropy production takes place
unless $S$ decays before the  Hubble parameter becomes equal to $m_{3/2}$.

The inflaton $S$ can also decay into the axino pair, which is represented by the combination $(\psi-\bar\psi)/\sqrt{2}$,
where $\psi$ and $\bar\psi$ denote the fermionic component of $\Psi$ and $\bar\Psi$, respectively.
This process, however, cannot be used as a reheating process because the produced axinos are not thermalized.

Therefore, the inflaton is required to decay soon after inflation in the present scenario.  
The inflaton can decay through the following superpotential:
\be
	W = kS Y \bar Y,   \label{kSYY}
\ee
where $Y (\bar Y)$ is the chiral superfield which has some gauge charges but no PQ charges
and $k$ is a coupling constant assumed to be real and positive.
For example, in the framework of the KSVZ axion model, it can be the MSSM Higgses,
$Y=H_u$ and $\bar Y = H_d$.\footnote{
	The inflaton can also decay into gluons through the heavy quark loops in the KSVZ model,
	but it is subdominant under the presence of the term (\ref{kSYY}).
}
This term also generates the sizable $\mu$-term for $k\sim \kappa$ 
since, as noted earlier, the VEV of $S$ is given by $\langle S\rangle \sim m_{3/2}/\kappa$.
Then, the decay rate is given by 
$\sim 1/(8\pi) k^2 m_S$ and the reheating temperature after inflation, $T_R$, is estimated as 
\be
     T_R \sim 10^{11} \mathrm{GeV} 
     \left(\frac{\kappa}{10^{-3}}\right) ^{1/2}
     \left(\frac{k}{10^{-3}}\right)
     \left(\frac{f_a}{10^{15}\mathrm{GeV}}\right)^{1/2},
\ee
taking $m_S = \kappa f_a$ into account.\footnote{
	The introduction of the superpotential (\ref{kSYY}) may modify the inflaton dynamics significantly.
	First, if $k> \kappa$, the additional Coleman-Weinberg correction larger than the original one
	arises. Second, if $k<\kappa$, the inflation ends at the point where $Y (\bar Y)$
	becomes tachyonic. Here we assume $k\simeq \kappa$ so that the original inflationary dynamics
	 is not much affected.
	 \label{footnote_kSYY}
}
Thus the reheating temperature can be about  $10^{11}$~GeV for $k = \kappa =10^{-3}$ and 
$f_a=10^{15}$~GeV.
This is high enough to induce the saxion oscillation, as seen in the following.

\begin{figure}[t]
\centering
\includegraphics [width = 8.5cm, clip]{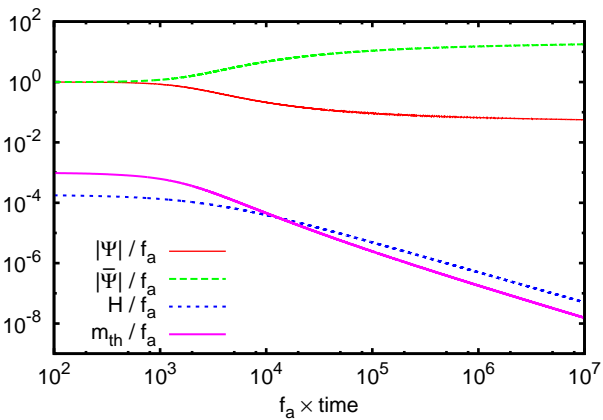}
\caption{
	Evolution  of the PQ fields, the Hubble parameter and the 
	thermal mass for $\kappa = 0.01$, $f_a = 0.01M_P$, and $T_R=0.1f_a$.
	The soft SUSY-breaking mass is not included.
}
\label{fig:PQ_dynamics2}
\end{figure}

One complexity arises from thermal effects on the scalar potential.
The reheating process thermalizes the Universe and the decay products of the inflaton 
interacts among others in the thermal bath.
In particular, in the KSVZ model, the $\Psi$ interacts with heavy quarks $Q$ and $\bar Q$, 
and they interact with MSSM particles through the QCD couplings.\footnote{
	In the DFSZ model, we also have to introduce such heavy quarks in order to solve the domain wall problem.
	This will be discussed in Sec.~6. Therefore, the presence of the finite-temperature potential 
	is generic for both the KSVZ and DFSZ models.
}
Then the two-loop effect induces a finite-temperature correction on the potential of $\Psi$
expressed as~\cite{Anisimov:2000wx}
\be
	V_\mathrm{th} \simeq 
	\alpha_s^2 T^4 \ln \frac{|\Psi|^2}{T^2}.	
	\label{thermal-log}
\ee
This thermal-log potential lifts up the flat direction,
and $|\Psi|$ ($|\bar\Psi|$) tends to roll down to a smaller (larger) value.
With use of the condition for the flat direction $\Psi \bar{\Psi} = f_a^2$, 
$|\bar\Psi|$ obtains an effective thermal mass from the thermal-log potential~(\ref{thermal-log}) as 
\be
	m_\mathrm{th}^2 \simeq \frac{\alpha_s^2 T^4}{|\bar\Psi|^2},
	\label{eq:thermal_mass}
\ee
and the equation of motion of $\bar\Psi$ is written as 
\be
	\ddot{\bar\Psi} + 3H \dot{\bar\Psi} 
	+ m_\mathrm{th}^2 \bar\Psi = 0. 	
	\label{eom_of_barPsi}
\ee
The evolution of $\bar{\Psi}$ is shown in Fig.~\ref{fig:PQ_dynamics2}.
Just after the reheating completes, both $\Psi$ and $\bar\Psi$ sits at $f_a$ as explained earlier,
and hence the thermal mass is larger than the Hubble parameter at this stage. 
Thus $\bar{\Psi}$ rolls down the thermal potential and 
$|\bar{\Psi}|$ increases until the effective mass~(\ref{eq:thermal_mass}) 
becomes comparable to the Hubble parameter.
Then the $\bar{\Psi}$ stops rolling and gets frozen. 
The frozen value of $|\bar\Psi|$ is determined at the time 
$H \sim m_\mathrm{th}$ and estimated as
\be
	|\bar\Psi| \sim \alpha_s M_P.   \label{Psi_finiteT}
\ee

\begin{figure}[t]
\centering
\includegraphics [width = 8.5cm, clip]{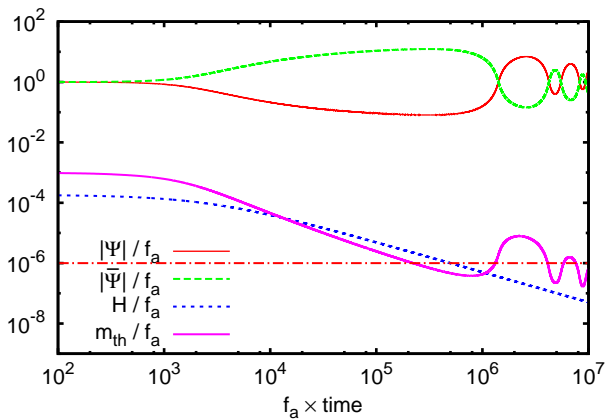}
\caption{Evolution  of the PQ fields, the Hubble parameter, and the 
	thermal mass for $m_{3/2} = 10^{-6}f_a$, 
	$\kappa = 0.01$, $f_a = 0.01M_P$, and $T_R=0.1f_a$.
}
\label{PQ_dynamics3_fig}
\end{figure}

After the temperature decreases and soft SUSY-breaking masses ($\sim m_{3/2}$) exceeds the thermal mass, 
the true minimum (\ref{vev_of_PQ}) appears.
On the other hand the PQ fields still remain frozen at (\ref{Psi_finiteT}) 
until the Hubble parameter becomes smaller than the soft SUSY-breaking masses.
Finally, after that, the PQ fields restart oscillation around 
its vacuum expectation value along the flat direction.
After all, the saxion begins to oscillate with an initial amplitude 
\be
   \sigma_i \sim \alpha_s M_P,
\ee
at $H\sim m_{3/2}$.
These features are confirmed by numerical calculation and seen in Fig.~\ref{PQ_dynamics3_fig}.
Although we have chosen specific parameters for a numerical reason in the figure,
qualitative arguments do not change for more realistic parameter choices.
We will see in the next section that the decay of the saxion 
dilutes the axion and its present cosmic density is acceptable.

\section{The late-time entropy production}

In this section, the cosmological scenario after the saxion begins to oscillate is described. 
As previously noted, due to the finite-temperature effect 
after reheating, the initial amplitude of the saxion oscillation 
is as large as  $\sigma_i \sim \alpha_s M_P$.
The saxion begins to oscillate when the Hubble parameter becomes
equal to the saxion mass, $m_\sigma$, in the radiation dominated Universe.
At that time, the ratio of the energy density of the saxion to the entropy density, 
which is a constant until the saxion decay, is calculated as
\be
	\begin{split}
	\frac{\rho_\sigma}{s} 
	&= \frac{1}{8} T_i \left( \frac{\sigma_i}{M_P} \right)^2 \\
	&\simeq 4 \times 10^6 \, \mathrm{GeV} 
	\left( \frac{m_\sigma}{1 \, \mathrm{TeV}} \right)^{1/2} 
	\left( \frac{\sigma_i}{\alpha_s M_P} \right)^2,
	\end{split}
\ee
where the subscript $i$ represents the value at the beginning of 
the saxion oscillation and 
\be
	T_i = \left(\frac{90}{\pi^2 g_*}\right)^{1/4} \sqrt{M_P m_\sigma} 
	\simeq 2\times 10^{10} \, \mathrm{GeV} 
	\left( \frac{m_\sigma}{1 \, \mathrm{TeV}} \right)^{1/2}.
\ee
Here we have used the value $g_* = 228.75$ as the relativistic degrees of freedom.
Since the saxion coherent oscillation behaves as a matter, 
the energy density of the saxion soon dominates the Universe if the initial amplitude is sufficiently large.

Here we estimate the saxion decay temperature.
From the saxion decay rate given previously
(\ref{eq:decay_rate_KSVZ}), the saxion decay temperature is estimated as
\be
	T_\sigma \simeq 5\, \mathrm{MeV}~ 
	\left( \frac{m_\sigma}{10\, \mathrm{TeV}} \right)^{3/2} 
	\left( \frac{10^{15} \, \mathrm{GeV}}{f_a} \right)
\ee
for the KSVZ axion model.
In order not to destroy the success of the Big Bang nucleosynthesis 
(BBN), the condition $T_\sigma \gtrsim {\rm a~few}\, \mathrm{MeV}$ must 
be imposed~\cite{Kawasaki:1999na}.
Here we have assumed that the decay into axions are suppressed.
As noted earlier, this is actually the case if $c_1 \simeq c_2$.\footnote{
 	The nonthermal axions produced by the saxion
	decay contributes to the extra radiation of the universe.
	If the branching ratio into the axions is around 0.1,
	it may account for the recently claimed existence of the extra radiation~\cite{Izotov:2010ca,Komatsu:2010fb,Dunkley:2010ge}.
}
Note also that rather large saxion mass of $\sim \mathcal(10{\rm TeV})$ is required
for successful reheating.
If the MSSM particle masses are around 1TeV, the saxion also decays into them with a similar decay rate
to the SM particles.
This is problematic since the lightest SUSY particles (LSP) are nonthermally produced
and its abundance is too large to be consistent with the WMAP observation, if the LSP is stable.
Thus we need to either introduce an $R$-parity violation in order for the LSP to decay rapidly~\cite{Hasenkamp:2011xh},
or assume all SUSY particles are so heavy that the saxion cannot decay into them kinematically.

In the case of DFSZ axion model, from the decay rate~(\ref{decay_rate_DFSZ}), we obtain
\be
	T_\sigma \simeq 5\, \mathrm{MeV} ~
	\left( \frac{m_\sigma}{1 \, \mathrm{TeV}} \right)^{3/2} 
	\left( \frac{10^{15} \, \mathrm{GeV}}{f_a} \right)
	\left( \frac{\mu}{m_\sigma} \right)^2.
\ee
In this case $T_\sigma \gtrsim 1$~MeV is satisfied 
for $f_a \simeq 10^{15} \, \mathrm{GeV}$ and  $m_\sigma \simeq 1 \, \mathrm{TeV}$.
Therefore, the saxion decay into SUSY particles can naturally be forbidden for TeV scale SUSY masses.

We can then calculate the dilution factor defined as the ratio of 
the entropy density before and after the saxion decay as
\begin{align}
	\gamma &= \frac{s_\mathrm{before}}{s_\mathrm{after}} 
	= \frac{3}{4} T_\sigma 
	\bigg( \frac{\rho_\sigma(t_i)}{s_i} \bigg)^{-1} \notag \\[1mm]
	&\simeq 2 \times 10^{-10} 
	\bigg( \frac{T_\sigma}{1 \, \mathrm{MeV}} \bigg) 
	\bigg( \frac{1 \, \mathrm{TeV}}{m_\sigma} \bigg)^{1/2}
	\bigg( \frac{\alpha_s M_P}{\sigma_i} \bigg)^2.
	\label{dilution_factor}
\end{align}

\subsection{The axion abundance}

Now let us estimate the axion abundance under the entropy production after the saxion decay.
The axion starts to oscillate when $H(T) = m_a(T)$, where $m_a(T)$ is the 
temperature-dependent axion mass given by~\cite{Gross:1980br,Turner:1985si}
\be
	m_a(T) \simeq 
	\begin{cases}
		0.08 m_a (\Lambda_\mathrm{QCD} / T )^{3.7} 
		& \text{ for } T > \Lambda_\mathrm{QCD} / \pi \\
		m_a & \text{ for } T < \Lambda_\mathrm{QCD}/\pi,
	\end{cases}
\ee
and $m_a$ is the zero temperature mass of the axion given by
\be
	m_a \simeq 6 \times 10^{-9} \, \mathrm{eV} 
	\bigg( \frac{10^{15} \, \mathrm{GeV}}{f_a} \bigg).
\ee
The energy density of the axion is given by $\rho_a = (1/2)m_a^2f_a^2 \theta_1^2$, 
where $\theta_1$ is the initial misalignment angle.
Here and hereafter the subscript ``1'' refers to the time when
the axion starts to oscillate.
Because the PQ symmetry is broken at the end of the inflation, 
the initial misalignment angle takes random values in each horizon.
Therefore, we take the averaged initial misalignment angle as $\pi / \sqrt 3$.
The ratio of energy density of the axion to that of the saxion 
at the beginning of the axion oscillation is given by
\be
	\frac{\rho_a (t_1)}{\rho_\sigma(t_1)} 
	= \frac{\pi^2 f_a^2}{18 M_P^2}.
\ee
Then the axion density to entropy ratio is
\be
	\frac{\rho_a(t_1)}{s_1} 
	= \frac{\pi^2 f_a^2}{18 M_P^2} \frac{\rho_\sigma(t_1)}{s_1}.
\ee
Using~Eq.(\ref{dilution_factor}), the present axion density
to entropy ratio is written as 
\be
	\frac{\rho_a(t_0)}{s_0} 
	= \gamma \xi^{-1}(T_1) \frac{\rho_a (t_1)}{s_1} 
	= \frac{\pi^2 f_a^2 T_\sigma}{24 M_P^2 \xi(T_1)},
\ee
where $\xi$ is defined as the ratio of the temperature-dependent
mass to zero temperature mass of the axion :
$\xi(T_1) \equiv m_a(T_1) / m_a$.
For the PQ scale $f_a$ as large as $10^{15}$GeV, the finite-temperature effect is negligible at 
$t = t_1$ and hence it is approximated as $\xi(T_1) \approx 1$~\cite{Kawasaki:1995vt}.
To show this, using the scale factor dependence of the energy density 
of radiation $\rho_r \propto a^{-3/2}$
when the saxion decays into radiation gradually in the saxion-dominated Universe, 
the relation
\be
	\frac{\rho_\sigma(t_\sigma)}{\rho_\sigma(t_1)} 
	= \bigg( \frac{T_\sigma}{T_1} \bigg)^8
\ee
is derived.
Using $\rho_\sigma(t_\sigma) = 3 \Gamma_\sigma^2 M_P^2$, 
$\rho_\sigma(t_1) = 3m_a^2 (T_1) M_P^2$, 
we get 
\be
	\frac{T_\sigma}{T_1} 
	= \bigg( \frac{\Gamma_\sigma}{m_a(T_1)} \bigg)^{1/4},
\ee
and hence the temperature that the axion begins to oscillate is 
estimated as
\be
	T_1 \simeq 0.2 \, \mathrm{GeV} ~
	\bigg( \frac{10^{15} \, \mathrm{GeV}}{f_a} \bigg)^{0.13} 
	\bigg( \frac{T_\sigma}{1 \, \mathrm{MeV}} \bigg)^{0.26}.
\ee
Thus, for $f_a\simeq 10^{15}$~GeV, the axion starts to oscillate 
below the QCD scale and then the finite-temperature effect can be 
negligible.

Finally, making use of the present ratio of the critical density to 
the entropy density $\rho_{\mathrm{cr},0}/s_0 \simeq 3.64 
\times 10^{-9} h^2 \, \mathrm{GeV}$,
the density parameter of the axion is calculated as
\be
	\Omega_a h^2 \simeq 0.02 ~ \xi^{-1}(T_1) 
	\bigg( \frac{T_\sigma}{1\, \mathrm{MeV}} \bigg) 
	\bigg( \frac{f_a}{10^{15} \, \mathrm{GeV}} \bigg)^2.
	\label{density_parameter_axion}
\ee
Therefore, the axion can be the dominant component of the DM
by taking account of the entropy production process from the saxion decay.
As described later, the axionic strings emit the axion and its contribution is comparable to the
coherent oscillation one.

\subsection{The gravitino and axino abundance}

It is well-known that in a SUSY theory, gravitinos are efficiently produced at the reheating
and they may be cosmologically harmful depending on the reheating temperature.
In the present model, however, the late-time entropy production dilutes the harmful relics.
Let us see it.

Gravitinos are produced both thermally~\cite{Bolz:2000fu} and nonthermally~\cite{Kawasaki:2006gs}
from the inflaton decay, but diluted sufficiently.
The thermally produced gravitino abundance, in terms of the number to entropy ratio, is estimated as
\begin{equation}
\begin{split}
	Y_{3/2}^{\rm (TP)} \simeq & 4\times 10^{-22}
	\left (\frac{T_{\rm R}}{10^{11}{\rm GeV}} \right )
	\left ( \frac{T_\sigma}{1{\rm MeV}}  \right ) \\[1mm]
	& \times \left( \frac{1{\rm TeV}}{m_\sigma} \right)^{1/2} 
	\left( \frac{\alpha_s M_P}{\sigma_i} \right)^2.
\end{split}
\end{equation}
The nonthermally produced one through the inflaton decay is given by
\begin{equation}
\begin{split}
	Y_{3/2}^{\rm (NTP)} \simeq & 4\times 10^{-27} 
	\left (\frac{10^{11}{\rm GeV}}{T_{\rm R}} \right )
	\left ( \frac{T_\sigma}{1{\rm MeV}}  \right )
	\left( \frac{1{\rm TeV}}{m_\sigma} \right)^{1/2} \\[1mm]
	&\times \left ( \frac{f_a}{10^{15}{\rm GeV}}  \right )^4
	\left ( \frac{\kappa}{10^{-3}}  \right )^2
	\left( \frac{\alpha_s M_P}{\sigma_i} \right)^2.
\end{split}
\end{equation}
These satisfy the bound on the unstable gravitino abundance from BBN 
$m_{3/2}Y_{3/2}\lesssim 10^{-13}$-$10^{-9}$GeV 
for $m_{3/2}\sim 1$-10~TeV~\cite{Kawasaki:2004yh,Jedamzik:2006xz}.

The axino, which is the fermionic superpartner of the axion, might also have
significant effects on cosmology once they are produced 
in the early universe~\cite{Rajagopal:1990yx,Goto:1991gq,Covi:2001nw,Cheung:2011mg}.
The axino abundance from thermal production~\cite{Covi:2001nw}, after the dilution, is estimated as
\begin{equation}
\begin{split}
	Y_{\tilde a}^{\rm (TP)} \simeq  & 1\times 10^{-19}
	\left( \frac{1{\rm TeV}}{m_\sigma} \right)^{1/2}
	 \left( \frac{T_{\rm R}}{10^{11}{\rm GeV}} \right) 
	\left( \frac{T_\sigma}{1{\rm MeV}} \right) \\[1mm]
	& \times \left( \frac{10^{15}{\rm GeV}}{f_a} \right)^2
	\left( \frac{\alpha_s M_P }{\sigma_i} \right)^2.   \label{axino}
\end{split}
\end{equation}
In the present model, the axino mass comes from the VEV of $S$.
It generates the axino mass of $m_{\tilde a}=\kappa \langle S\rangle \sim m_{3/2}$.
Thus the axino mass is comparable to the gravitino.
If the axino is not the LSP, it has a similar lifetime to the saxion in the KSVZ model, 
and it decays before BBN. The constraint is given as $Y_{\tilde a} \lesssim 10^{-12}$
so as not to produce too much LSPs.
If the axino is the LSP, the bound reads $m_{\tilde a}Y_{\tilde a} \lesssim 4\times 10^{-10}$GeV.
In both cases, the constraint is satisfied as is seen in Eq.~(\ref{axino}).

A more important contribution to the axino comes from the direct decay of the inflaton $S$.
As noted in Sec.~\ref{sec:post}, the $S$ decays into the axino pair.
The branching ratio into the axino pair, $B_{\tilde a}$, is evaluated as $B_{\tilde a} \sim \kappa^2/(4k^2)$.
The axino abundance produced in this way is written as
\begin{equation}
\begin{split}
	Y_{\tilde a}^{\rm (NTP)} \simeq  & 1\times 10^{-12} B_{\tilde a}
	\left( \frac{1{\rm TeV}}{m_\sigma} \right)^{1/2}
	 \left( \frac{T_{\rm R}}{10^{11}{\rm GeV}} \right) \\
	& \times \left( \frac{T_\sigma}{1{\rm MeV}} \right) 
 	\left( \frac{10^{15}{\rm GeV}}{f_a} \right)
	\left( \frac{10^{-3}}{\kappa} \right)
	\left( \frac{\alpha_s M_P }{\sigma_i} \right)^2.   \label{axino}
\end{split}
\end{equation}
This is close to the observational upper limits shown above.
\footnote{
	The branching ratio $B_{\tilde a}$ can be suppressed if $k > \kappa$ is allowed.
	However, the condition $k > \kappa$ affects the inflationary dynamics as shown in footnote \ref{footnote_kSYY}.
	In this case, the larger $f_a$ is derived from the CMB normalization, which is not desirable.
	}

Therefore, the cosmological problems related to the gravitino and axino are solved
due to the late-time entropy production from the saxion decay.

\section{Axionic strings and domain walls}

After the PQ phase transition, the $U(1)_{\rm PQ}$ symmetry is spontaneously broken,
and axionic strings are formed accordingly through the Kibble mechanism.
The explicit breaking of the $U(1)_{\rm PQ}$ by the anomaly effect further breaks it down to
a $Z_N$ symmetry, where $N$ is the color anomaly number.
At the QCD phase transition, $Z_N$ symmetry is spontaneously broken and domain walls are produced.
Therefore, a complicated network of topological defects appears such that
each string is attached by $N$ domain walls.
In particular, domain walls are harmful because they soon come to dominate the Universe.

In the case of $N=1$, however, there is no such a domain wall problem.
The reason is following.
At formation, the energy of such a domain wall-string system is dominated 
by the string but soon the energy of the domain wall becomes dominant. 
Once domain walls dominate the wall-string system, 
the surface area tends to become smaller by the surface tension of the domain walls.
As a result, the domain wall shrinks and disappears. 
Thus  the cosmological problem does not arise~\cite{Vilenkin:1982ks}.
Through these processes, axionic strings and walls lose their energy emitting axions.
The emitted axion obtains a mass after the QCD 
phase transition and gives significant contributions to the present 
energy density of the Universe in addition to the coherent oscillation.
According to Refs.~\cite{Yamaguchi:1998gx,Hiramatsu:2010yu}, 
the energy density of the axion radiated from strings is comparable to that of the coherent oscillation.
This is true even in the presence of late-time entropy production,
since both the coherent oscillation and string-induced axion are diluted in a similar way.
In the KSVZ model with one set of heavy quarks, the color anomaly number is equal to 1
and no domain wall problem arises.

In the case of $N \geq 2$, however, the produced domain walls have 
complicated structure. $N$ sheets of walls are attached to each string. 
This wall-string system dominates the Universe 
immediately and the present energy density of the domain walls becomes 
much larger than the critical energy density of the Universe.
It is a cosmological disaster.
This is the case for the DFSZ axion model.
One of the solutions to the problem is to introduce some heavy quarks like the KSVZ axion model
in order to make the color anomaly number 1 ~\cite{Georgi:1982ph}.
Actually these heavy quarks, introduced to solve the domain wall problem, 
serve as a source for the finite-temperature potential which induces the saxion oscillation 
with large amplitude, as discussed earlier.

\section{Baryogenesis through the Affleck-Dine mechanism}

Because all contents of the Universe are diluted by the late-time 
entropy production, we need large baryon asymmetry enough to survive 
the dilution.
Such a large initial baryon asymmetry can be generated by considering 
the Affleck-Dine (AD) mechanism~\cite{Affleck:1984fy,Dine:1995kz}.
Actually the AD mechanism may work well 
even if the late-time entropy production dilutes the baryon number~\cite{Kawasaki:2007yy}.
The MSSM contains many scalar 
fields which are scalar superpartners of the SM fermions, 
and there exist flat directions in the scalar potential
which do not feel the potential in the SUSY limit at the renormalizable level.
Such a flat direction is called the AD field and is parametrized by $\phi$.
The potential of the AD field is lifted up by the nonrenormalizable superpotential 
\be
    W_\mathrm{NR} = \frac{\phi^n}{nM^{n-3}} \quad 
    \text{with} \quad n \geq 4,    \label{WNR}
\ee
where $M$ is some cutoff scale.
Then the potential for the AD field is written as 
\be
\begin{split}
   V_S(\phi) &= m_\phi^2 |\phi|^2 - c_H H^2 |\phi|^2 \\
   & + \bigg( a_m m_{3/2} \frac{\phi^n}{nM^{n-3}} 
   + \mathrm{h.c.} \bigg) + \frac{ |\phi|^{2(n-1)}}{M^{2(n-3)}}
\end{split}
\ee
where $m_\phi$ is the mass of the AD field which is of order of
the gravitino mass $m_{3/2}$ in the gravity-mediation scenario
and $a_m$ is a complex-valued coefficient of order unity. 
Here we have also included the Hubble-induced mass term with real-valued 
$O(1)$ coefficient $c_H$~\cite{Dine:1995kz}, which is assumed to be positive here.
In addition, the finite-temperature correction is significant 
as noted in the previous section. This is represented as
\be
   V_\mathrm{th}(\phi) 
   = \sum _{f_k |\phi| < T} c_k f_k^2 |\phi|^2 T^2 
   + a \alpha_s(T)^2 T^4 \ln \bigg( \frac{|\phi|^2}{T^2} \bigg),
\ee
where $c_k$ and $a$ is real and positive coefficient of order unity 
and $f_k$ is an appropriate coupling constant.
Thus the potential of the AD field is given by
\be
   V(\phi) = V_S(\phi)  + V_\mathrm{th}(\phi).
\ee
In the early Universe, the Hubble parameter is large and 
the AD field sits at the temporal minimum determined by the balance between the Hubble mass
term and the nonrenormalizable term,
\be
    |\phi| \simeq \big( H M^{n-3} \big)^{1/(n-2)}.
\ee
The value of $|\phi|$ changes adiabatically as $H$ decreases.
The AD field begins the coherent oscillation when the Hubble parameter 
falls below the value $H_\mathrm{osc}$ defined as
\be
    H_\mathrm{osc}^2 \equiv m_\phi^2 
    + \sum_{f_k |\phi| < T} c_k f_k^2 T^2 
    + a \alpha_s(T)^2 \frac{T^4}{|\phi|^2}.
\ee

In our scenario, it is assumed that the oscillation of the AD 
field is caused by the thermal-log potential, i.e. 
$H_\mathrm{osc}^2 \simeq \alpha_s^2 T^4 / |\phi|^2$,
and the oscillation starts before the decay of the inflaton, 
$H_\mathrm{osc} > \Gamma_I$, where $\Gamma_I$ is the decay rate of the inflaton.
In this period, the temperature of the Universe is given by
$T \approx (T_R^2 H M_P)^{1/4}$.
Since the resultant baryon asymmetry is extremely small 
in the case of $n=4$~\cite{Kawasaki:2007yy}, we consider the case of $n=6$.
The baryon number density at the time when the AD field starts 
to oscillate is $n_B(t_{\rm osc}) \approx \delta_\mathrm{CP} 
m_{3/2} |\phi_\mathrm{osc}|^2$,
where $\delta_\mathrm{CP}$ is an effective CP phase which is assumed to be $O(1)$.
After the reheating caused by the inflaton decay, the saxion starts to oscillate at $t=t_i$.
Since the saxion produces the present entropy in our model,  
the baryon-to-entropy ratio is fixed at that time. 
Using $|\phi_\mathrm{osc}| \simeq (H_\mathrm{osc} M^3)^{1/4}$ 
and $H_\mathrm{osc} \simeq (\alpha_s^2 T_R^2 M_P M^{-3/2} )^{2/3}$,
the baryon-to-entropy ratio after the decay of the saxion is estimated as
\begin{widetext}
\begin{align}
    \frac{n_B}{s} =& \frac{3T_\sigma}{4} 
    \frac{n_B(t_i)}{\rho_\sigma(t_i)}  
    = \frac{3T_\sigma}{4} 
    \frac{n_B(t_{\rm osc})}{\rho_\sigma(t_i)}
    \left(\frac{a(t_{\rm osc})}{a(t_i)}\right)^3
    = \frac{3T_\sigma}{4} 
    \frac{n_B(t_{\rm osc})}{\rho_\sigma(t_i)}
    \left(\frac{\Gamma_I}{H_{\rm osc}}\right)^2 
    \left(\frac{m_{3/2}}{\Gamma_I}\right)^{3/2} \notag \\[0.6em]
    \simeq & 3 \times 10^{-10}\delta_\mathrm{CP}
    \left( \frac{T_\sigma}{1 \, \mathrm{MeV}} \right) 
    \left( \frac{m_{3/2}}{1 \, \mathrm{TeV}} \right)^{1/2} 
    \notag 
     \left( \frac{10^{11} \, \mathrm{GeV}}{T_R} \right)
     \left( \frac{\alpha_s M_P}{\sigma_i} \right)^2 
     \left(\frac{M}{100 M_P} \right)^3.
\end{align}
\end{widetext}
Hence, the desired baryon asymmetry, $n_B / s \sim 10^{-10}$, is generated
by choosing the cutoff scale $M$ appropriately.
Notice that the baryonic isocurvature perturbation is sufficiently small and satisfies the observational bound~\cite{Komatsu:2010fb}.

Let us check the consistency of the AD mechanism with present inflation model.
During inflation, the AD field has VEV of $\sim (H_I M^3)^{1/4}$,
where $H_I = \kappa f_a^2 /\sqrt{3}M_P$ is the Hubble scale during inflation.
Thus the superpotential (\ref{WNR}) induces an effective constant term given by
$W_{\rm eff} \sim (H_I M)^{3/2}/6$.
Numerically, this is comparable to the constant term $W_0 = m_{3/2}M_P^2$
for the above parameter choices.
Thus the inflaton dynamics is not affected by the presence of the large amplitude of the AD field.

\section{A variant model : smooth-hybrid inflation from a SUSY axion model}

So far we have analyzed the SUSY axion model based on the superpotential 
(\ref{susy_axion_superpot}) and shown that it causes hybrid inflation.
In this section we consider a variant type of the SUSY axion model.

Let us take the superpotential 
\begin{equation}
	W = S\left( \frac{(\Psi \bar\Psi)^{n}}{M^{2(n-1)}}  - \mu^2  \right) + \lambda \Psi X\bar X,
	\label{smooth}
\end{equation}
where $M$ is a cutoff scale and $n \geq 2$.
In addition to the PQ symmetry and $R$-symmetry whose charges are given in Table~\ref{charge_asignments},
this superpotential also has a discrete symmetry $Z_{n}$ under which $\bar\Psi$ has a charge $+1$
and others have zero.
This has a flat direction along $\Psi\bar\Psi = (\mu M^{n-1})^{2/n}$,
and it is stabilized by the SUSY-breaking masses similar to the model of (\ref{susy_axion_superpot}).
Thus the minimum of the potential lies at $|\Psi|\sim |\bar\Psi| \sim f_a \equiv (\mu M^{n-1})^{1/n}$,
where the PQ symmetry is spontaneously broken with scale of $f_a$. 

The superpotential (\ref{smooth}), which provides one of the SUSY axion models,
coincides with that causing a so-called smooth-hybrid inflation~\cite{Lazarides:1995vr,Yamaguchi:2004tn}.
Although the post-inflationary saxion dynamics is similar to the previous model,
there is a significant difference between hybrid and smooth-hybrid inflation.
In the smooth-hybrid inflation model, the $\Psi (\bar\Psi)$ has a nonzero VEV during inflation,
hence the PQ symmetry is already broken.
Therefore, no topological defects are formed after inflation and we do not need to worry about
the possibly harmful domain wall problem.
The scalar spectral index is predicted to be around $0.97$ without introducing nonminimal 
K\"ahler potentials.

Instead, since the PQ symmetry is broken and the axion obtains a quantum fluctuation 
during inflation, it may generate a large cold dark matter (CDM) isocurvature perturbation.
The amplitude of the CDM isocurvature perturbation is given by
\begin{equation}
	S_{\rm m}\simeq \frac{\Omega_a}{\Omega_{\rm m}} \frac{H_I}{\pi \psi_N \theta_i},
\end{equation}
where $\Omega_{\rm m}$ is the DM density parameter
and $\psi_N$ denotes the field value of the PQ scalar 
when the cosmological scales exit the horizon.
The axion density parameter is evaluated in a way similar to (\ref{density_parameter_axion}), 
but in the present case the initial misalignment angle $\theta_i$ can be chosen arbitrarily
since the PQ symmetry is already broken during inflation and a region with angle $\theta_i$ 
is expanded to cover the whole observable region of the Universe.
Thus we have
\be
	\Omega_a h^2 \simeq 0.03 ~ \xi^{-1}(T_1)
	\bigg( \frac{T_\sigma}{1\, \mathrm{MeV}} \bigg) 
	\bigg( \frac{f_a}{10^{15} \, \mathrm{GeV}} \bigg)^2
	 \theta_i^2.
	\label{axion_smooth}
\ee
By demanding that the correct magnitude of the density perturbation
and the scalar spectral index of $\simeq 0.97$ are reproduced,
we find $\mu \sim 3\times 10^{14}{\rm GeV}$ and $M\sim 10^{15}$GeV for $n=2$.
The situation does not change much for $n>2$.
Thus we obtain the PQ scale as $f_a \sim 5\times 10^{14}$GeV.
Numerically, $\psi_N$ is slightly smaller than $f_a$.
The inflation scale is calculated as $H_I = \mu^2 /(\sqrt 3 M_P) \sim 10^{10}$GeV.
Substituting these parameters, we obtain $S_{m} \sim 10^{-6}$ for $\theta_i \simeq  0.1$
which is close to the observational bound on the isocurvature perturbation~\cite{Komatsu:2010fb}.
In this case the axion cannot be the dominant component of DM.

\section{Conclusions}

We have shown that an inflation naturally takes place in the framework a SUSY axion model.
Identifying the PQ scalar fields as the waterfall fields, the hybrid inflation is realized.
In this case, the PQ phase transition takes place at the end of inflation.
Note that the isocurvature perturbation does not arise 
since the PQ symmetry is restored during inflation.
In order for the inflation to account for the observed density perturbation, 
the PQ symmetry-breaking scale must be around $10^{15}$GeV.
In addition, the observed value of the spectral index can be reproduced by introducing 
a nonminimal K\"ahler potential.
Considering the post-inflationary dynamics, we have found that 
the saxion begins to oscillate with large initial amplitude and its decay produces a huge entropy.
Thanks to this late-time entropy production process, the axion is diluted and 
its present density becomes consistent with that of DM.
Simultaneously, the gravitinos~\cite{Bolz:2000fu,Kawasaki:2006gs}
and axinos~\cite{Covi:2001nw} are also diluted by 
the late-time entropy production, and hence the cosmological problems of gravitino or 
axino do not  arise in the present scenario~\cite{Kawasaki:2010gv, Kawasaki:2008jc}.
On the other hand, the preexisting  baryon asymmetry is also 
diluted, so we need an initial baryon asymmetry large enough to survive the dilution.
We have pointed out that the desired baryon asymmetry is obtained through the AD mechanism.
Finally, topological defects formed at the PQ phase transition 
or the QCD phase transition disappear without causing any cosmological problems
if the color anomaly number is equal to 1.
Therefore, the SUSY axion model studied in this paper not only
provides a solution to the strong CP and the hierarchy problems 
but also accounts for inflation and the DM of the Universe.

We have also shown that a slight modification of the SUSY axion model 
results in a so-called smooth-hybrid inflation.
In this case no topological defects are formed and the scalar spectral index 
is consistent with the WMAP observation naturally.
Instead, it predicts a large isocurvature perturbation close to the present observational bound.
Or it may be discovered by the future cosmological measurements.

\section*{Acknowledgment}

K.N. would like to thank Fuminobu Takahashi for useful discussion.
This work is supported by Grant-in-Aid for Scientific research from
the Ministry of Education, Science, Sports, and Culture (MEXT), Japan,
No.\ 14102004 (M.K.), No.\ 21111006 (M.K. and K.N.), No.\ 22244030 (K.N.) and also 
by World Premier International Research Center
Initiative (WPI Initiative), MEXT, Japan.

  

\end{document}